\documentclass[twocolumn,preprintnumbers,amsmath,amssymb,floatfix]{revtex4}
\usepackage{graphicx}
\usepackage{dcolumn}
\usepackage{bm}

\begin{document}
\title{Atmospheric leptons\\
the search for a prompt component}
%
%

\author{Thomas K. Gaisser\\
Bartol Research Institute and Department of Physics and Astronomy\\
University of Delaware, Newark, DE USA
          }

\begin{abstract}%
The flux of high-energy ($\ge$GeV) neutrinos consists primarily 
of those produced by
cosmic-ray interactions in the atmosphere.
The contribution from extraterrestrial sources is still
unknown.  Current limits suggest that the observed
spectrum is dominated by atmospheric neutrinos up to  at least
$100$~TeV.
The contribution of charmed hadrons to the flux of atmospheric
neutrinos is important in the context of the search for astrophysical
neutrinos because the spectrum of such	``prompt" neutrinos
is harder than that of ``conventional" neutrinos from decay
of pions and kaons.  The prompt component therefore
becomes increasingly important as energy increases.  
This paper reviews the
status of the search for prompt muons and neutrinos with emphasis on
the complementary aspects of muons, electron neutrinos and muon neutrinos.
\end{abstract}
\maketitle
\section{Introduction}
\label{sec1}
Before experimental discovery of charmed hadrons at accelerators
in the mid-1970s
there was intense interest in using atmospheric muons
to find evidence for production of heavy, short-lived hadrons.
For example, a highlight of the 1973 cosmic-ray conference at
Denver was an update of measurements made over several years with
the underground muon spectrometer at Park City, Utah~\cite{ICRC73}.
At first, the observed angular dependence of multi-TeV
muons had appeared to be more isotropic than could be explained
solely by production through the pion and kaon channels, which
are strongly enhanced at large zenith angles.  With an
improved understanding of the overburden and the detector response, however,
it was finally concluded that the Park City data were consistent
a ``conventional" origin from decay of pions and kaons.  An isotropic
``prompt" component was not manifest for energies below $10$~TeV.

Production of charmed hadrons has now been measured over a large
range of energy at accelerators.  The production cross section increases 
significantly from approximately $\sim1\,\mu$b at
$\sqrt{s} \approx 10$~GeV to several~mb at $\sqrt{s} \approx 7$~TeV~\cite{sigmaCharm}.
There is still not full coverage of phase space for charm production, however.
In particular, the level of "intrinsic" charm~\cite{Brodsky} 
production is still uncertain.  The SELEX measurement~\cite{SELEX} shows
a large asymmetry in the ratio of charm to anti-charm baryons
produced by baryon beams on a fixed target, while little
or no asymmetry is observed in a pion beam.  This observation indicates
some level of intrinsic charm in which the valence quarks of the
projectile pick up a charmed quark.  Charmed hadrons
produced as fragments of the incident nucleon beam will contribute
disproportionately to the spectrum of atmospheric leptons because
of the steep cosmic-ray energy spectrum.  Thus, even if intrinsic
charm contributes less to the total cross section for producing charm 
than production via QCD processes, it may
have a significant effect on the prompt contribution to atmospheric
muons and neutrinos.

In addition to the intrinsic interest in identifying the charm
contribution to the fluxes of atmospheric muons and neutrinos,
there is another,  perhaps more important, reason for trying
to measure this component.  That is because of the relevance of prompt
neutrinos in the search for neutrinos of astrophysical origin.
Like a flux from unresolved extra-galactic neutrino sources,
the prompt contribution is isotropic for $E_\nu < 10^7$~GeV.
It is also harder by one power of energy
than the spectrum of conventional atmospheric
neutrinos.  For these reasons, prompt neutrinos constitute an
important background for neutrino astronomy.

The paper begins in \S~\ref{sec2} with a discussion of the ingredients
needed to calculate fluxes of atmospheric muons and neutrinos, including
relevant analytic approximations and the primary cosmic-ray spectrum.
Section~\ref{sec3} reviews models for charm production and
corresponding predictions for fluxes of muons and neutrinos.
In \S~\ref{sec4} we calculate the fluxes of conventional atmospheric
muons and neutrinos and compare them with the charm contribution.
The effect of the knee of the primary spectrum is included.
The predictions are illustrated in \S~\ref{sec5}
with approximate calculations of the event rates
for detector with a gigaton target volume like IceCube~\cite{IceCube1,IceCube2,IceCube3}.
The concluding Section~\ref{sec6} comments on the current status
and prospects for detection of prompt leptons in the near future.

\begin{figure*}
\centering
\includegraphics[width=8cm]{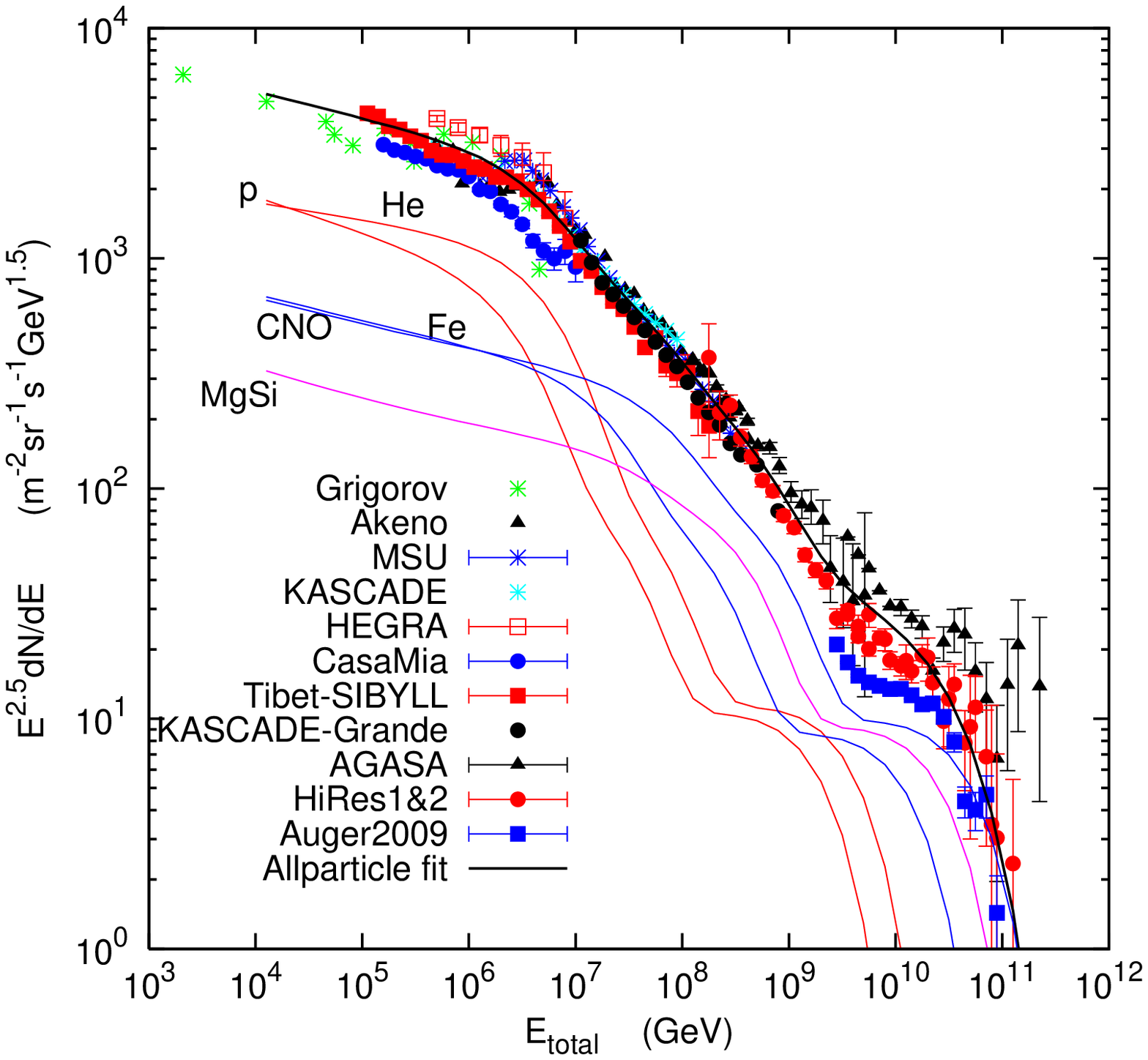}\,\includegraphics[width=7.cm]{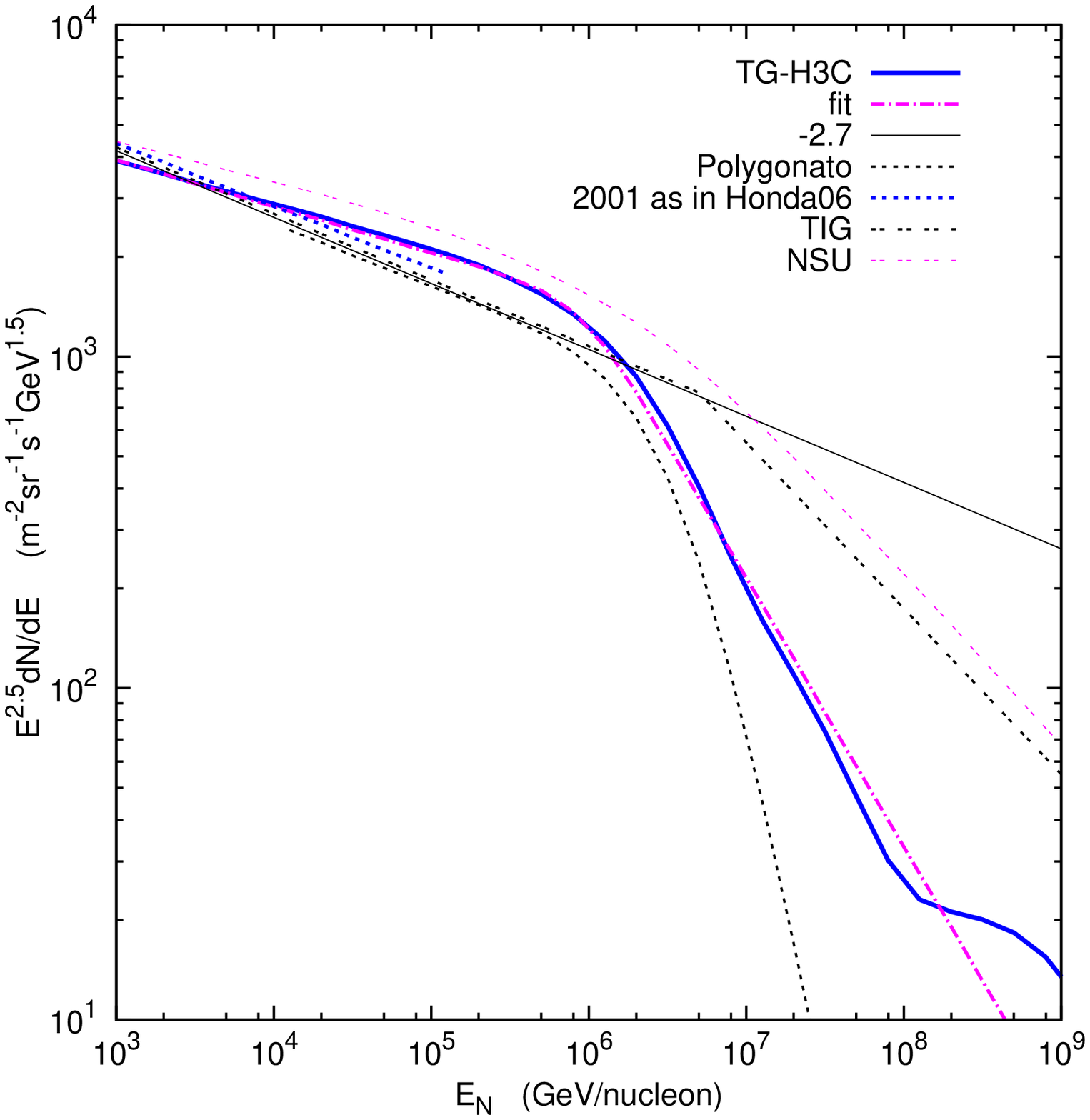}
\caption{Left: All-particle spectrum from Ref.~\cite{Gaisser2012} where references to the data are given.
Right: Spectrum of nucleons for several assumptions (see text for explanation). }
\label{fig1}       
\end{figure*}

\section{Atmospheric muons and neutrinos}
\label{sec2}
The two main ingredients in the calculation of atmospheric neutrinos
are the primary spectrum and the hadronic physics of meson production
in hadronic interactions.  Because production of pions, kaons and charmed hadrons 
occurs at the nucleon level, what is most relevant is the primary
spectrum of nucleons per GeV/nucleon.  Composition comes in through the
ratio of protons to total nucleons, which determines the charge ratio
of muons and particle/anti-article ratio for neutrinos.

\subsection{Primary spectrum}
For illustration I use a phenomenological model of the primary spectrum
with three populations of particles and five nuclear components~\cite{Gaisser2012},
as shown in Fig.~\ref{fig1} (Left).
There are two basic assumptions.  First, it is assumed
that all energy dependence (whether from
acceleration or propagation) depends only on
how particles are affected by their magnetic environment.  As a
consequence, each nuclear component (mass number $Z$ and total momentum per
particle $P$) 
depends on magnetic rigidity ($R$) in the same way, where $R=Pc/Ze$.
Peters~\cite{Peters}
pointed out the consequence of this assumption for the primary composition
in the region of the knee of the spectrum, namely, that,
when expressed in terms of total energy per particle, protons would steepen
first followed by helium and then by nuclei with successively higher charge.

The other assumption, following Hillas~\cite{Hillas}, is that three populations
of particles are sufficient to characterize the entire cosmic-ray spectrum.
This is almost certainly an oversimplification.  A more realistic picture
would likely involve many individual sources injecting particles at
various distances and times, as in the model for galactic cosmic rays of Blasi and Amato~\cite{Blasi}.  
Thus the three populations represent three classes of sources:
\begin{enumerate}
\item Particles accelerated by supernova remnants in the galaxy, 
\item A higher energy galactic component of uncertain origin, and
\item Particles accelerated at extra-galactic sources.
\end{enumerate}
The contribution of nuclei of mass $A_i$ to the all-particle
spectrum is given by
\begin{equation}
\phi_i(E)\;=\;\frac{E{\rm d}N_i}{{\rm d}E}\;=\;\Sigma_{j=1}^3\,a_{i,j}\,
E^{-\gamma_{i,j}}\times \exp\left[-{E\over Z_i R_{c,j}}\right],
\label{ModelH3}
\end{equation}
where $E$ is the total energy per nucleus.

\begin{table*}
\centering
\begin{tabular}{r|r|ccccc}  \hline
$R_c$ & $\gamma$&  p & He & CNO & Mg-Si & Fe \\ \hline  \hline
$\gamma$ for Pop. 1 & ---- & 1.66 & 1.58 & 1.63 & 1.67 & 1.63 \\ \hline
Population 1: $ 4$ PV & see line 1 & 7860  &   3550   &   2200     &   1430 & 2120  \\ \hline
Pop. 2: $30$ PV & 1.4 &20  &   20     &   13.4     &   13.4 & 13.4  \\
Pop. 3 (mixed): $2$ EV &1.4 & 1.7  &   1.7     &   1.14     &   1.14 & 1.14  \\ 
    " (proton only): $60$ EV & 1.6 & 200. & 0 & 0 & 0 & 0 
\end{tabular}
\caption{Cutoffs, integral spectral indices and normalizations constants $a_{i,j}$
for Eq.~\ref{ModelH3}.}
\label{tab1}
\end{table*}

The spectral indices for each group and the normalizations are given explicitly in
Table~\ref{tab1}.  The parameters for Population 1 are based on fits to spectra
of nuclear groups measured by CREAM~\cite{CREAM,CREAM1},
which we assume can be extrapolated to a rigidity of $4$~PV to describe the knee.  This
is an unverified simplifying assumption that needs to be checked by measurements in
the PeV range.
In Eq.~\ref{ModelH3} $\phi_i$ is d$N/$d$\ln E$ and $\gamma_i$ is the integral spectral index.
The subscript $i=1,5$ runs over the standard five groups (p, He, CNO, Mg-Si and Fe), and the 
all-particle spectrum is the sum of the five.

The spectrum of nucleons as a function of energy per nucleon corresponding to Eq.~\ref{ModelH3} is
given by
\begin{equation}
\phi_N(E_N)\;=\;\frac{E{\rm d}N}{{\rm d}E_N}\;=\;\Sigma_{i=1}^5\,A_i\times\phi_i(A\,E_N).
\label{EperN}
\end{equation}
Because of the steep cosmic-ray spectrum, protons are relatively more important and heavy nuclei 
less important in the spectrum of 
nucleons (as a function of $E_N=E_{\rm tot}/A$) than in the all particle spectrum.

The spectrum of nucleons is plotted for several assumptions in Fig.~\ref{fig1} (right).
 The straight solid line shows a simple $E^{-2.7}$ spectrum of nucleons to guide the eye.
The straight dotted line shows the spectrum below 100~TeV 
recommended in 2001~\cite{GHLS} as a standard
for use in calculating fluxes of atmospheric leptons up to 10 TeV. 
At low energy the fit was based on measurements of BESS~\cite{BESS98} and AMS~\cite{AMS}.  The
spectral index used for protons at the time was rather steep (2.74), based on the measurements of
BESS and AMS below 200 GeV.  Recent results of PAMELA~\cite{PAMELA} show that the spectrum
of protons hardens above 200 GeV.  Two options
were given for helium, which contributes of order 25\% of the spectrum of nucleons.  The high
option for helium (with an integral spectral index of $\gamma=1.64$) 
suggested by emulsion chamber measurements in the 10 TeV range at the
time~\cite{RUNJOB,JACEE} has since been
confirmed by ATIC~\cite{ATIC}, CREAM~\cite{CREAM} and PAMELA~\cite{PAMELA}.  
A version of the spectrum of Ref.~{GHLS} is used in the standard
calculations of the flux of atmospheric neutrinos by Honda et al.~\cite{Honda2} and by the Bartol
group~\cite{Barr}.   
The spectrum of Honda et al. (as described in~\cite{Honda1})  
uses a harder spectrum for hydrogen (2.71 instead of 2.74)
above 100 GeV.  Their overall spectrum is nearly constant at $\gamma=1.69$ from 200 GeV to 50 TeV
with a fraction of helium that increases from 20\% to 25\% in the same region.
The spectral index of the
spectrum of nucleons in the model of Ref.~\cite{Gaisser2012} is nearly constant at $\gamma=1.63$
from 200 GeV to 50 TeV with a corresponding increase in the contribution of helium
from 22\% to 30\%.  The contribution from heavier nuclei is at the level of 10\%.

The other lines in the spectrum of nucleons all include the effect of the knee in the cosmic-ray
spectrum in one way or another.  
The heavy solid curve is the nucleon spectrum corresponding to the model in Table~\ref{tab1}.
The nearby pink dash-dot is an analytic approximation to that model, which is described
below in Eq.~\ref{approx-spectrum}.  The strong knee around 1~PeV is the consequence of
the increasing fraction of heavy nuclei in the model.
In addition to the model of Ref~\cite{Gaisser2012},
Fig.~\ref{fig1} also shows the polygonato model~\cite{Polygonato} without any contribution
from nuclei heavier than iron.  Each nuclear component in the model steepens by
$\delta=1.9$ at a rigidity of $4.49$~PV.
The effect of the knee begins to show up in the nucleon spectrum already somewhat below
one PeV.  Using the rule of thumb that there is on average a factor of ten between the parent cosmic
ray energy and the secondary leptons, taking account of the steepening of the spectrum will
be important for muon and neutrino energies of 100 TeV and above, which we is discussed in
\S~\ref{sec4}.  The double dotted line that steepens from a differential index of $-2.7$ to
$-3.0$ at $5\times 10^6$~GeV is the primary spectrum used in the charm calculation
of Ref.~{TIG}.

\subsection{Hadron production and decay kinematics in the atmosphere}

The phenomenology of atmospheric leptons depends on the production of pions, kaons and
heavier hadrons by interactions of cosmic-rays in the atmosphere and on the kinematics
for the relevant decay channels.  Production and subsequent decay occur through
generation by a steep spectrum of primary and secondary cosmic rays in the atmosphere.
The competition between reinteraction and decay of unstable hadrons depends on density
and altitude.  In the framework of a set of analytic approximations for solution of
the cascade equations, the essential dependence on energy and 
zenith angle comes through the critical energy defined as
\begin{equation}
E_{\rm crit}\,=\,\frac{\epsilon_\iota}{\cos\theta^*}\,=\,\frac{m_\iota c^2 h_0}{c\,\tau_\iota},
\end{equation}
where the index $\iota$ indicates the hadron ($\pi^\pm$, $K^\pm$, $K_L$ or charmed hadron),
$\tau_\iota$ is the meson lifetime, $h_0$ is the scale height of the atmosphere
and $\theta^*$ is the zenith angle ($^*$corrected for curvature of the Earth for $\theta \ge 70^\circ$).
Values of the important characteristic energies are given in Table~\ref{tab2}.
\begin{table}
\centering
\begin{tabular}{l|cccc} 
Particle ($\alpha$): & $\pi^\pm$&  $K^\pm$ & $K_L^0$ & Charm  \\   \hline
$\epsilon_\alpha$ (GeV):  & 115 & 850 & 205 & $\sim 3\times 10^7$
\end{tabular}
\caption{Characteristic energies.}
\label{tab2}
\end{table}

For a power-law
spectrum of primary nucleons, the expression for the lepton
spectrum factorizes into a product of the primary spectrum and an
expression that reflects the properties of production of secondary
hadrons by the cosmic-ray spectrum and their subsequent decay to muons and 
neutrinos.
\begin{eqnarray}
\phi_\nu(E_\nu)& =  &\phi_N(E_\nu) \times
 \left\{{A_{\pi\nu}\over 1 + 
B_{\pi\nu}\cos\theta\, E_\nu / \epsilon_\pi}\right. \nonumber \\
& & \left.+\,\,\,{A_{K\nu}\over 1+B_{K\nu}\cos\theta\, E_\nu / \epsilon_K}\right.\nonumber \\
& & \left. +\,\,\,{A_{{\rm charm}\,\nu}\over 1+B_{{\rm charm}\,\nu}\cos\theta\, E_\nu / \epsilon_{\rm charm}}\right\},
\label{angular}
\end{eqnarray}

The $A$-factors in Eq.~\ref{angular} are a product of the spectrum weighted 
moments for production of mesons by nucleons times the spectrum weighed
moments of the meson decay distributions, which include both the decay
kinematics and the branching ratios.  For a power-law spectrum of
decaying pions with a differential spectral index $\alpha$ the decay factor is
\begin{equation} 
{1-r_\pi^{\alpha}\over\alpha\,(1-r_\pi)}\;\;{\rm and}\;\;\;
{(1-r_\pi)^\alpha\over\alpha\,(1-r_\pi)}
\end{equation}
for muons and neutrinos respectively.
In the low energy limit, the spectral index is the same as that of the primary spectrum of nucleons,
but at high energy the spectrum of the decaying pions is one power steeper because of the competition
between decay and reinteraction.
Low and high are with respect to the critical energy $\epsilon_\pi/\cos(\theta)$.
The ratio $r_\pi = m_\mu^2/m_\pi^2 = 0.5731$.  

The forms for two-body decay
of charged kaons are the same, but the mass ratio factor is much smaller: $r_K=0.0458$.  
The larger critical energy for charged kaons
leads to an increase in the contribution of kaons with increasing energy
for both muons and neutrinos.
The differences between the kinematic factors for two-body decay to
neutrinos and muons amplifies the importance of the kaon channel for neutrinos.
At high energy, in the TeV range and above, charged kaons account for about 80\%
of muon neutrinos as compared to 25\% of muons.

Each term in Eq.~\ref{angular} is a form that combines the low energy and high energy
solutions to the cascade equation respectively for pions, kaons and charmed hadrons in
the atmosphere.  The numerator is a product of the spectrum weighted moment for meson
production and the spectrum weighted moment of the decay distribution to $\nu_\mu$~\cite{Lipari93}
with $\alpha=\gamma+1$, where $\gamma$ is the integral spectral index of the
spectrum of primary nucleons.  The denominator governs the transition between the low
and the high energy regimes.
The forms for muons are similar.  For low energy, meson decay dominates over reinteraction
and the resulting lepton spectrum has the same shape as the primary spectrum of nucleons.
Charmed hadrons are in the low-energy regime for $E_{\rm lepton}< 10^7$~GeV.
For high energy ($E_{\rm lepton}>\epsilon_\alpha/\cos\theta$), reinteraction of the hadron
is more likely and the lepton spectrum becomes one power of energy steeper than the primary
spectrum.  

In the high energy limit, the spectrum weighted moment for meson decay has to be
evaluated on the steeper spectrum, and the attenuation lengths 
for reinteraction come into play.  The $B_{ij}$ quantities
in the denominators are the product of the ratios of low-energy to high energy
decay distributions combined with the function of attenuation lengths that accounts
for cascading of the mesons~\cite{Gaisser1990}.  
Explicitly, for neutrinos
\begin{equation}
B_{\pi\nu}\,=\,\left({\gamma + 2\over \gamma+1}\right)\,\left({1\over 1-r_\pi}\right)\,
\left({\Lambda_\pi - \Lambda_N\over \Lambda_\pi\ln(\Lambda_\pi/\Lambda_N)}\right)
\label{Bnufactor}
\end{equation}
and for muons
\begin{equation}
B_{\pi\mu}\,=\, \left({\gamma + 2\over \gamma+1}\right)\,\left({1-(r_\pi)^{\gamma+1}\over 1-(r_\pi)^{\gamma+2}}\right)\,
\left({\Lambda_\pi - \Lambda_N\over \Lambda_\pi\ln(\Lambda_\pi/\Lambda_N)}\right).
\label{Bfactor}
\end{equation}
The forms for kaons are the same as functions of $r_K$ and $\Lambda_K$.
The dependence of $\gamma$ on energy in the case of a non-power law primary
spectrum needs to be accounted for.

For a power-law primary spectrum of nucleons and assuming Feynman scaling for hadron production,
the cascade equations can be solved analytically as in Eq.~\ref{angular}.  The primary spectrum
can always be described locally as a power law, and similarly the hadronic interactions
can be written in terms of the scaled energy ($x=E_\alpha/E_N$) for a given primary energy per nucleon.
In both cases the dependence on energy is sufficiently gradual that the approximate analytical forms
can be used for quantitative calculations if
the slow variation with energy is accounted for.
This approach is taken in the calculation of Thunman, Ingelmann and Gondolo (TIG)~\cite{TIG}, 
which I follow here.   They define energy-dependent Z-factors 
as in the following example for nucleons producing charged pions:
\begin{equation}
Z_{N\pi^\pm}(E)\,=\,\int_E^\infty\,{\rm d}E'\frac{\phi_N(E')}{\phi_N(E)}
\frac{\lambda_N(E)}{\lambda_N(E')}\frac{{\rm d}n_{\pi^\pm}(E',E)}{{\rm d}E}.
\label{TIG-Z}
\end{equation}
Here $\lambda_N(E)$ is the nucleon interaction length and d$n_\pi^\pm$ is the
number of charged pions produced in d$E$ by nucleons of energy $E'$, and $\phi_N(E)$
is the spectrum of nucleons.  The energy-dependent Z-factors are then used
in applicable version of Eq.~\ref{angular} to evaluate the lepton spectrum.
This approximation is valid to the extent that the energy dependences are gradual.
They showed that the numerical approximation
based on the spectrum weighted moments taken from the interaction model
used in their Monte Carlo produced similar results to the full Monte Carlo.  
The advantage is that the analytic approximations can be tuned to match well to a particular
model of hadronic interactions.  They can then be used to extend a Monte Carlo
calculation to arbitrarily high energies without the statistical problems
that arise from the fact that mesons usually interact rather than decay at high energy.
This method has been used recently
in Ref.~\cite{FBD} to complement Monte Carlo calculations of atmospheric leptons
in the region of the knee of the cosmic-ray spectrum.

\section{Expectations for prompt leptons}
\label{sec3}

Limits on the prompt contribution to atmospheric muons
are conventionally expressed in terms of the ratio $R_c$ of
charm production to pion production by rewriting Eq.~\ref{angular} as
\begin{eqnarray}
\phi_\mu(E_\nu)& =  &\phi_N(E_\mu) \times
 \left\{{A_{\pi\mu}\over 1 + 
B_{\pi\mu}\cos\theta\, E_\mu / \epsilon_\pi}\right. \nonumber \\
& & \left.+{A_{K\mu}\over 1+B_{K\mu}\cos\theta\, E_\mu / \epsilon_K}
+{A_{\pi\mu}\times R_c}\right\}.
\label{angular2}
\end{eqnarray}
This form applies for lepton energy $<\,10^7$~GeV where the charm contribution is isotropic.
The current upper limit from LVD~\cite{LVD} is $R_c\leq 2 x 10^{-3}$ assuming a differential
primary spectrum $\propto E^{-2.77}$.  The primary spectrum of nucleons used in this
analysis is less steep by about $0.1$, so the LVD limit at $10$~TeV (for example) would be
reduced by a factor of two for comparison with the models as discussed here.  In what 
follows I will compare the current models for prompt leptons to $R_c\simeq 10^{-3}$
considered as an experimental upper limit.  

\begin{figure}
\centering
\includegraphics[width=8cm]{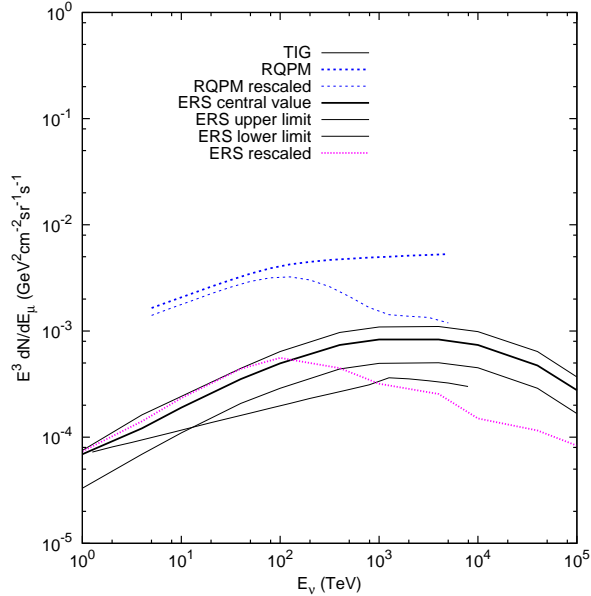}
\caption{Predictions of three models for the flux of
prompt muons.  See text for discussion of the rescaled plots.}
\label{fig2}       
\end{figure}

\subsection{Calculations of charm production}

Reference~\cite{TIG} uses a simple Monte Carlo to generate
the distribution of hadronic interactions and decays in the atmosphere and Pythia~\cite{Pythia} to
generate the secondary hadrons at each interaction point.  Charm production is calculated
within Pythia using first order QCD matrix elements to calculate $c\bar{c}$ production by
gluons and by quarks.  A renormalization factor $K=2$ is used to represent higher order QCD effects.
In addition to their Monte Carlo calculation, 
They parameterized their results
for the charm contribution in a form similar to the third term of Eq.~\ref{angular} as
\begin{equation}
\phi_C(E)\;=\;N_0\frac{E^{-(\gamma+1)}}{1\,+\,AE}
\label{TIGcharm}
\end{equation} 
with $A\approx 3\times 10^7$~GeV and $\gamma = 1.77$ below and $\approx 2.0$ above $10^6$~GeV. 
The value of $R_c$ for this model is $\simeq 2\times 10^{-5}$ at $10$~TeV.

The reference calculation used for evaluation of the atmospheric neutrino background in IceCube
at high energy is that of Enberg, Reno \& Sarcevic (ERS)~\cite{ERS}.  It is a QCD calculation
that gives results somewhat higher than some previous calculations~\cite{TIG,MRS}
and lower than an earlier NLO-QCD calculation~\cite{PRS}.  The ERS calculation
assumes the same primary spectrum of nucleons as in TIG~\cite{TIG}, which is shown in
the right panel of Fig.~\ref{fig1}.  The shape of the ERS calculation is similar to that of TIG,
and its central value is approximately a factor of two higher.  ERS assign an uncertainty
range of approximately $\pm50$\% to their calculation.  The value of 
$R_c$ for this model is $\simeq 10^{-4}$ at $10$~TeV.

The ``Recombination quark-parton model'' (RQPM)~\cite{RQPM} embodies the idea of
intrinsic charm.  The underlying concept is that--also for heavy quarks--there
is a process of associated production in which the $c\bar{c}$  pair produced
when the projectile proton fragments can recombine with valence quarks (di-quarks)
and with sea quarks to produce charmed hadrons, including charmed hyperons.
For example, in this model the process $p\rightarrow \Lambda_c^+ + \bar{D}^0$
would be expected at a level $(m_s/m_c)^2$ relative to associated production
of strangeness, $p\rightarrow \Lambda K^+$.  The parameters of the model
are adjusted to fit the then available data on charm production in Ref.~\cite{BMNSST},
and a parameterization of the prompt muon flux in the energy range from $5$~TeV to 
$5000$~TeV is given.
The value of 
$R_c$ for this model is $\simeq 8\times 10^{-4}$ at $10$~TeV.  The RQPM model
is close to the LVD upper limit, but still allowed by it.  It is interesting
that the recent IceCube limit
on the prompt contribution to $\nu_\mu$-induced upward muons~\cite{Anne}
is just at the level of the RQPM model.

Figure~\ref{fig2} compares the predictions of these three models for
the prompt contribution to the atmospheric muon flux.  An important point to
note in comparing the curves in this figure is that the spectrum also depends
on what is assumed for the primary spectrum.  The TIG and ERS use the same 
assumption as each other, in which the knee is probably too high in energy.
The primary spectrum used in Ref.~\cite{BMNSST} is taken from a
model~\cite{NSU} in which the knee is attributed to energy losses in
the sources (photo-disintegration for nuclei and photo-pion production
for nuclei.  In this case, the knee is a function of $E/A$ rather than
rigidity for the nuclei, and the knee occurs at higher energy per nucleon
for protons than for helium because of the relatively high threshold for
photo-pion production as compared to photo-disintegration.  
This spectrum also has a knee at rather
high energy and, in addition, appears to be anomalously high even in the few TeV range,
as shown in Fig.~\ref{fig1}.  The broken lines labeled ``rescaled'' in the figure
are estimates of what the ERS and RQMP models would give for the prompt muon
flux if the spectrum of Ref.~\cite{Gaisser2012} had been used.  This estimate is
obtained by multiplying by the ratio at $E_{\rm N} = 10\times E_\nu$ of the
nucleon flux used here~\cite{Gaisser2012} to those used for ERS~\cite{ERS} and
RQPM~\cite{BMNSST}.

\begin{figure*}
\centering
\includegraphics[width=7.5cm]{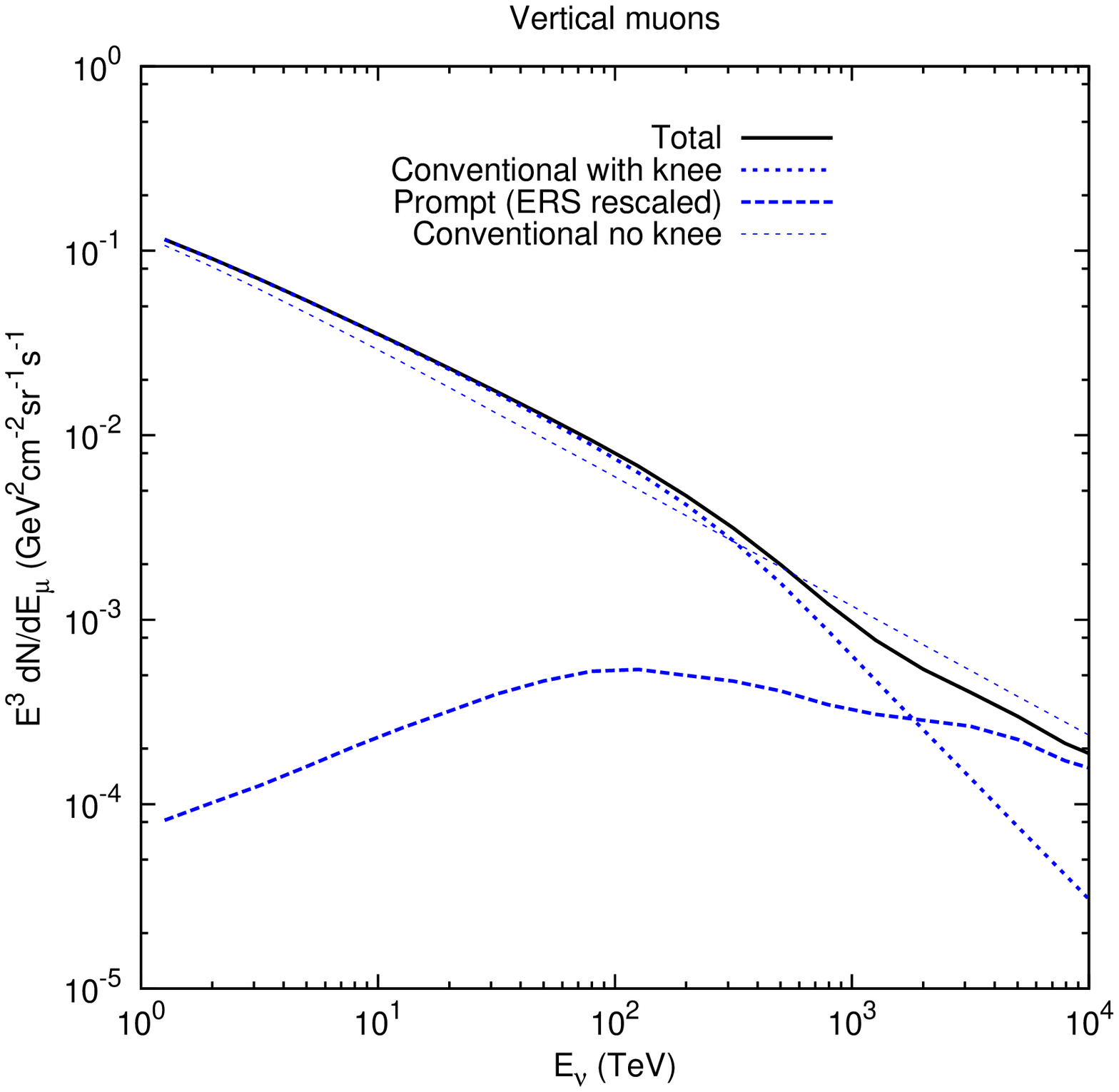}\,\includegraphics[width=7.5cm]{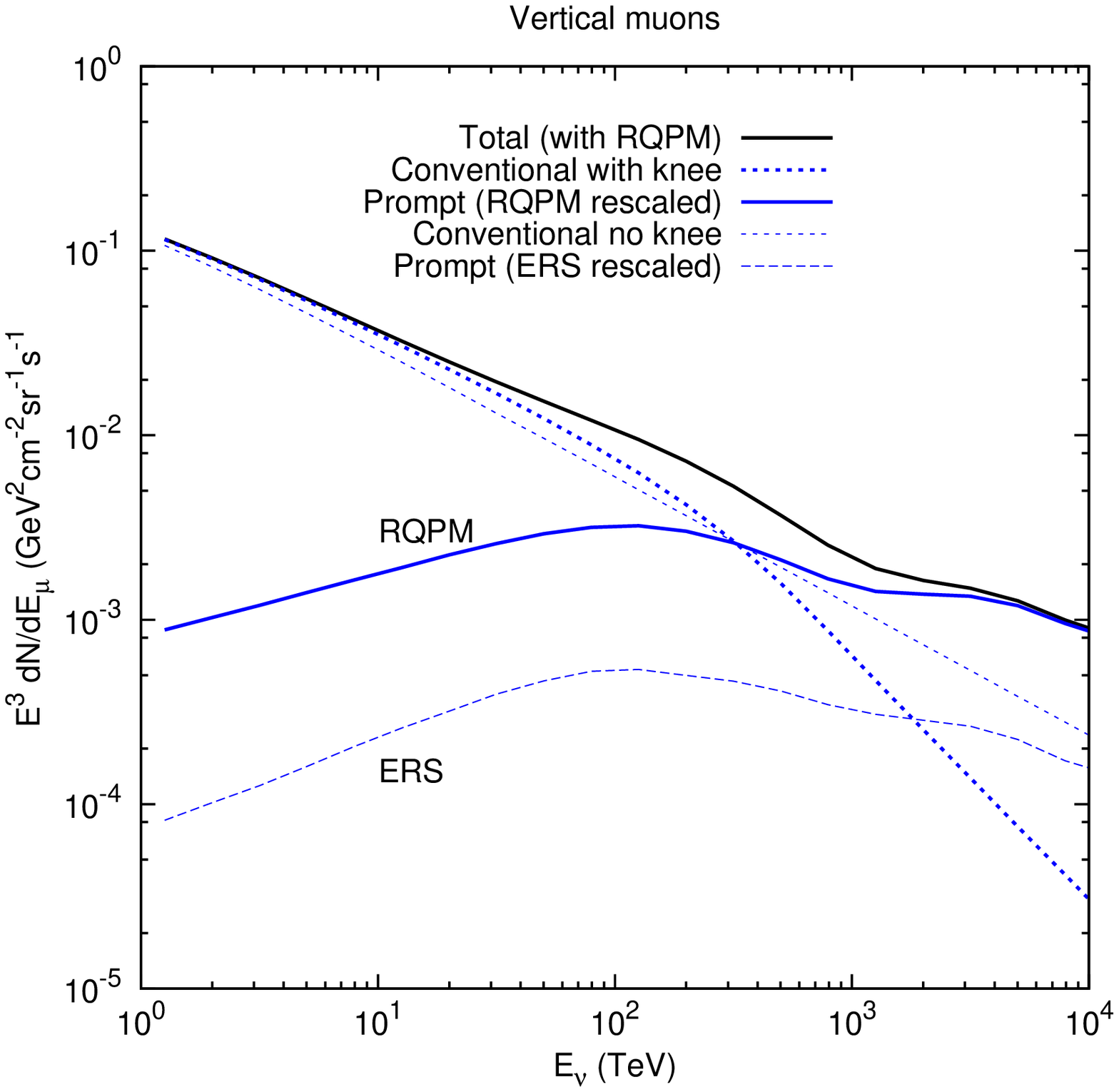}
\caption{Muon spectra including prompt contribution.  Left: prompt component from ERS model (rescaled);
Right: prompt component from RQPM model (rescaled).}
\label{fig3}       
\end{figure*}

\section{Atmospheric leptons including the knee}
\label{sec4}

My presentation at ISVHECRI-2012
showed a series of figures in which the contribution of
charm was shown separately from the ``conventional''
atmospheric leptons from decay of pions and kaons.  In those
figures, a simple power law spectrum was assumed for evaluating
the fluxes of the conventional leptons, and the charm contributions
were taken directly from the models.  Here I take account of the knee
in the spectrum, and I make a preliminary effort to present
a consistent representation of the prompt component by rescaling
their fluxes to the same spectrum of nucleons used for the
conventional leptons, as described in the previous paragraph.

Since the primary
  spectrum is no longer a power law, the spectrum-weighted moments
  depend on energy.  To evaluate energy-dependent Z-factors,
  Eq.~\ref{TIG-Z} is used.  For simplicity here
 only the energy dependence of the spectrum is considered, and  
standard values of nucleon interaction and attenuation lengths are used.
In addition, a scaling form for meson production is assumed.  The goal
here is to demonstrate the effect of the knee in the cosmic-ray spectrum
on the lepton fluxes.

\subsection{Approximation for pion and kaon production}

Explicit approximations for scale-independent meson production forms
are given in Ref.~\cite{Gaisser2002}.  The approximation for nucleons
to produce charged pions is
\begin{eqnarray}
F_{N\pi}(E_\pi/E_N)&=&E_\pi\frac{{\rm d}n_\pi(E_\pi/E_N)}{{\rm d}E_\pi}\\ \nonumber
&\approx&c_+(1-x)^{p_+}\,+\,c_-(1-x)^{p_-}
\label{pionapprox}
\end{eqnarray}
with a similar form for production of kaons.  With these scaling forms for particle
production, the integral
in Eq.~\ref{TIG-Z}
can be rewritten as
\begin{equation}
Z_{N\pi^\pm}\,=\,\int_0^1\,\frac{{\rm d}x}{x^2}\frac{\phi_N(E/x)}{\phi_N(E)}F_{N\pi^\pm}
\label{Tom}
\end{equation}

\begin{figure*}
\centering
\includegraphics[width=7.5cm]{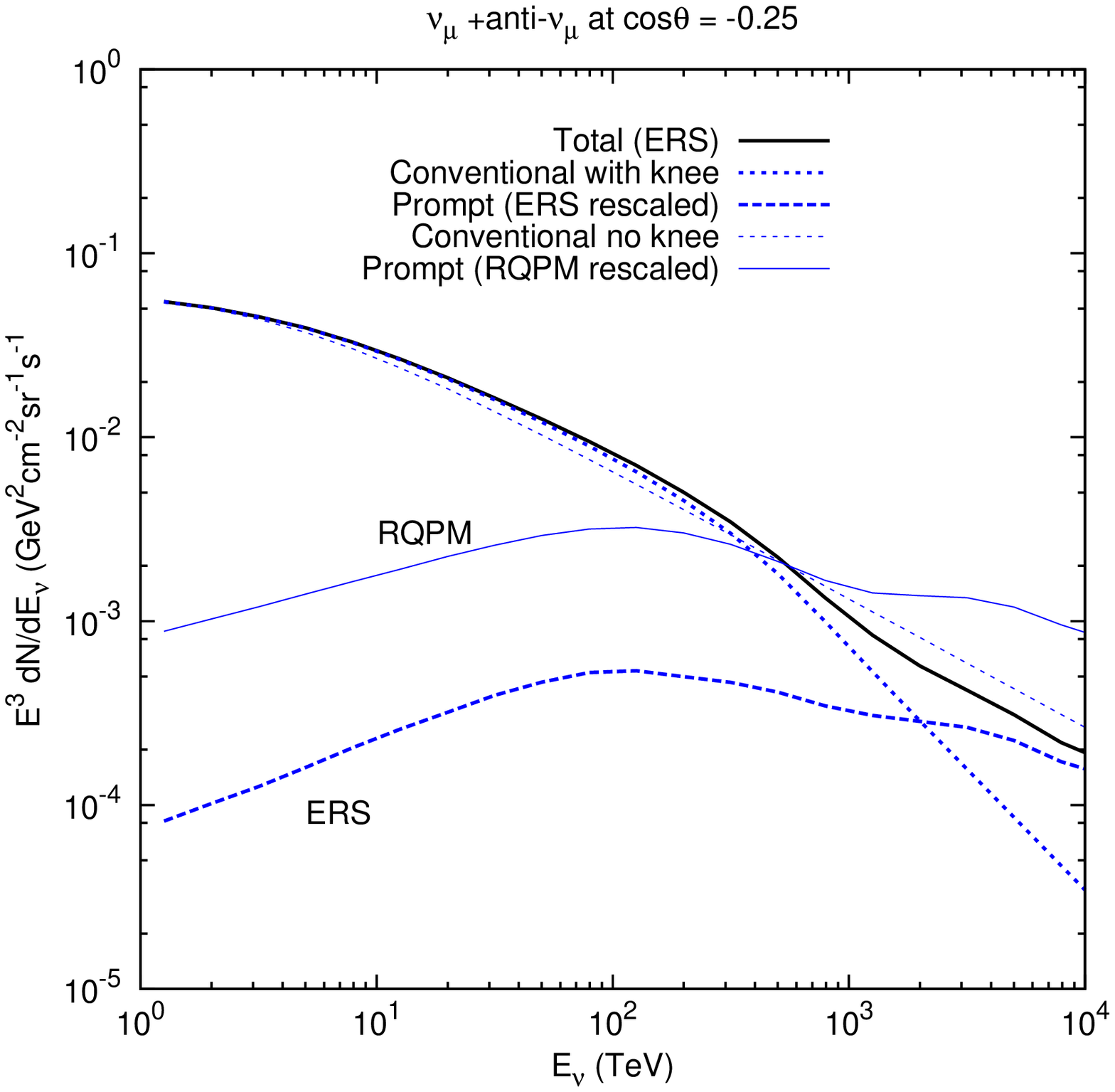}\,\includegraphics[width=7.5cm]{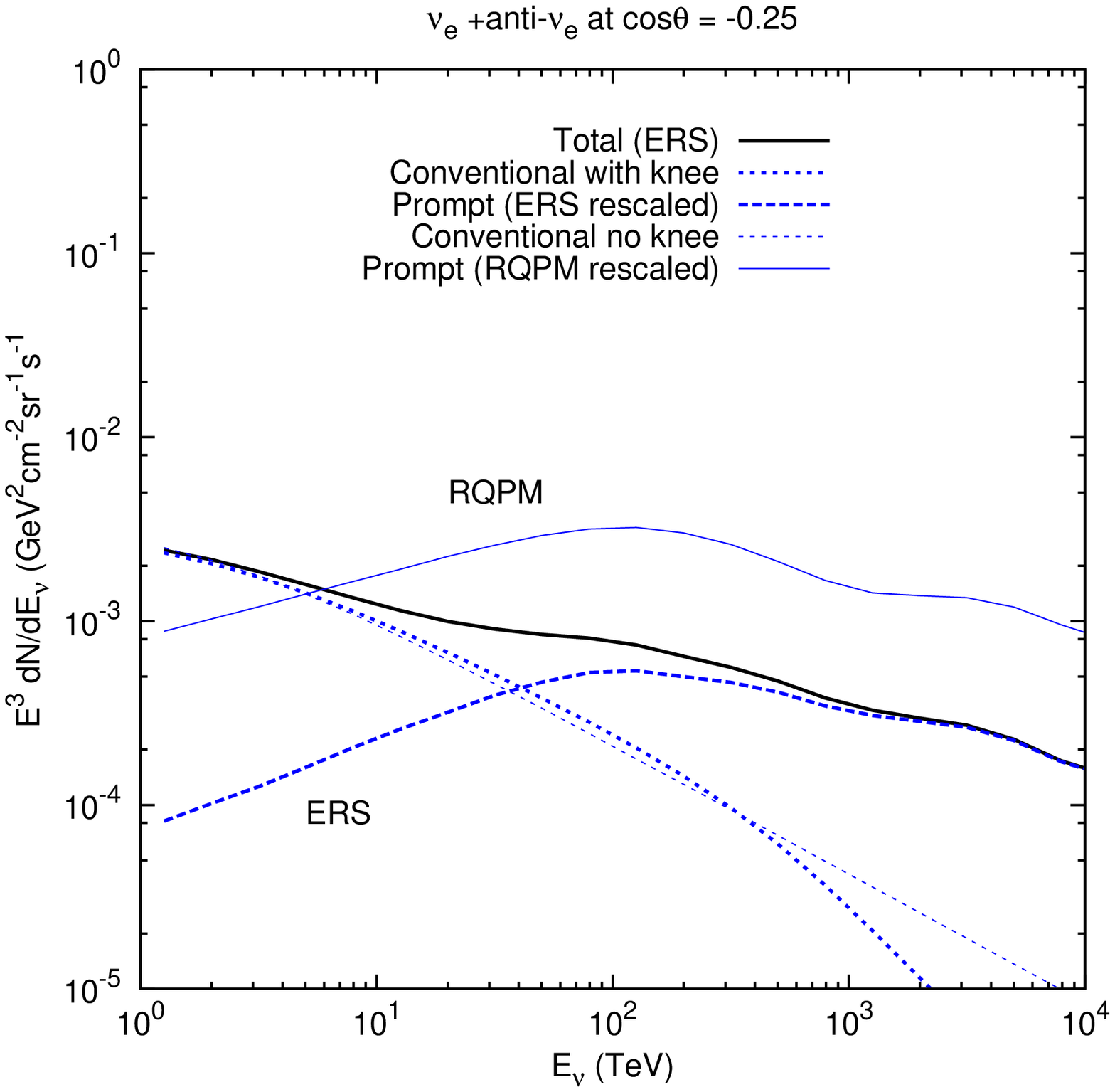}
\caption{Neutrino spectra including the prompt contribution.  Left: $\nu_\mu+\bar{\nu}_\mu$;
 Right: $\nu_e+\bar{\nu}_e$.}
\label{fig4}       
\end{figure*}

\subsection{Approximation for nucleon spectrum}

The nucleon spectrum with the knee as parameterized in Table~\ref{tab1} and Eq.~\ref{EperN}
can be approximated well (better than 10\% to 30 PeV) 
with the standard two-power-law form of Ref.~\cite{Samvel} to describe the knee, a form
which is also used
in Ref.~\cite{Polygonato}.  Specifically,
\begin{equation}
E\frac{{\rm d}N}{{\rm d}E}\,=\,const\times E^{-\gamma}\,(1\,+\,(E/E^*)^\epsilon)^{-\delta/\epsilon},
\label{approx-spectrum}
\end{equation}
with $\gamma = 1.64$, $\delta = 0.67$, $E^*=9.E+5$~GeV and $\epsilon=3.0$.
This approximation locates the knee in the nucleon spectrum just below a PeV, and
the slope steepens from an integral spectral index of $1.64$ below the knee to
$2.31$ after the knee.  The normalization constant is $10290.$~m$^{-2}$sr$^{-1}$s$^{-1}$.
The steepening in the nucleon spectrum is a consequence of the steepening of
the all-particle spectrum amplified by the increasing fraction of nuclei in the all-particle spectrum.
Equations~\ref{pionapprox}, \ref{Tom} and \ref{approx-spectrum} are combined
and integrated numerically to obtain energy-dependent Z-factors.

\subsection{Atmospheric leptons including the knee}
The fluxes of $\mu^\pm$, $\nu_\mu+\bar{\nu}_\mu$ from decay of pions and kaons are obtained using the 
energy-dependent Z-factors to evaluate Eq.~\ref{angular}.  The energy-dependence
of the spectral index in the meson decay factors that appear in Eq.~\ref{angular}
are also accounted for by using the local (energy-dependent) integral spectral index of the nucleon 
spectrum, which steepens gradually from $\gamma = 1.64$ to $2.31$ through the knee region.  

Calculation of the flux of electron neutrinos requires tracing the contributions of
Ke3 decays of both charged and neutral kaons.  This has been done by taking account
of the neutron/proton ratio of the primary nucleons and tracking separately 
the production of $K\bar{K}$ pairs and 
production of kaons in association with $\Lambda$ and $\Sigma$ hyperons.
An approximate value of $n/p\,=\,0.54$ has been used~\cite{Gaisser2012}.
As noted in the previous section, associated production of kaons by dissociation of an incident nucleon
into a kaon and a hyperon is a prototype for intrinsic charm.

Results for the lepton fluxes are shown in Figs.~\ref{fig3} and~\ref{fig4}.  The vertical muon fluxes
are shown in Fig.~\ref{fig3}, comparing the ERS model for prompt muons (left) to the RQPM model (right).
Below the knee the RQPM prompt flux is rescaled down by about 15\%, while the ERS prompt muon flux
is rescaled up by about 20\%.  The rescaled ERS prompt muons are repeated on the right panel for
comparison.  In the region between 100 TeV and 1 PeV the rescaled RQPM prompt flux is a factor
of 5 to 6 higher than the rescaled ERS flux.  The RQPM prompt component crosses the conventional
flux around 300 TeV, as compared to a crossover at 2~PeV for ERS (left panel).

Figure~\ref{fig4} compares the situation for $\nu_\mu$ (left panel) with that for $\nu_e$ (right panel).
The ERS model is shown for the prompt component.  The RQPM flux is also repeated on both plots for
reference.  The conventional atmospheric $\nu_e$ flux is approximately a factor 20 lower than
the conventional flux of $\nu_\mu$, so the electron neutrino component is dominated by the
prompt component at quite low energy.

\begin{figure*}
\centering
\includegraphics[width=7.5cm]{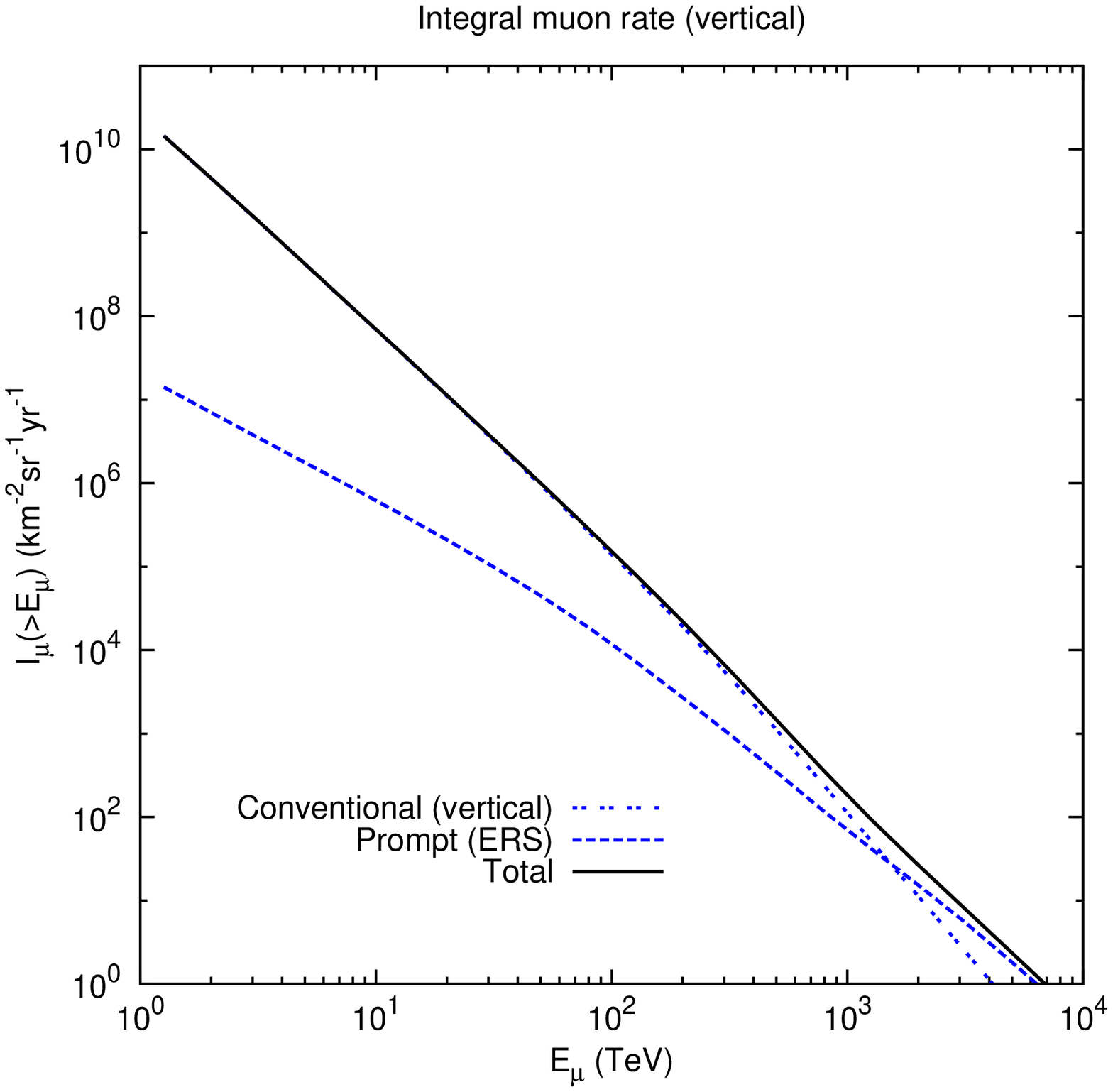}\,\includegraphics[width=7.5cm]{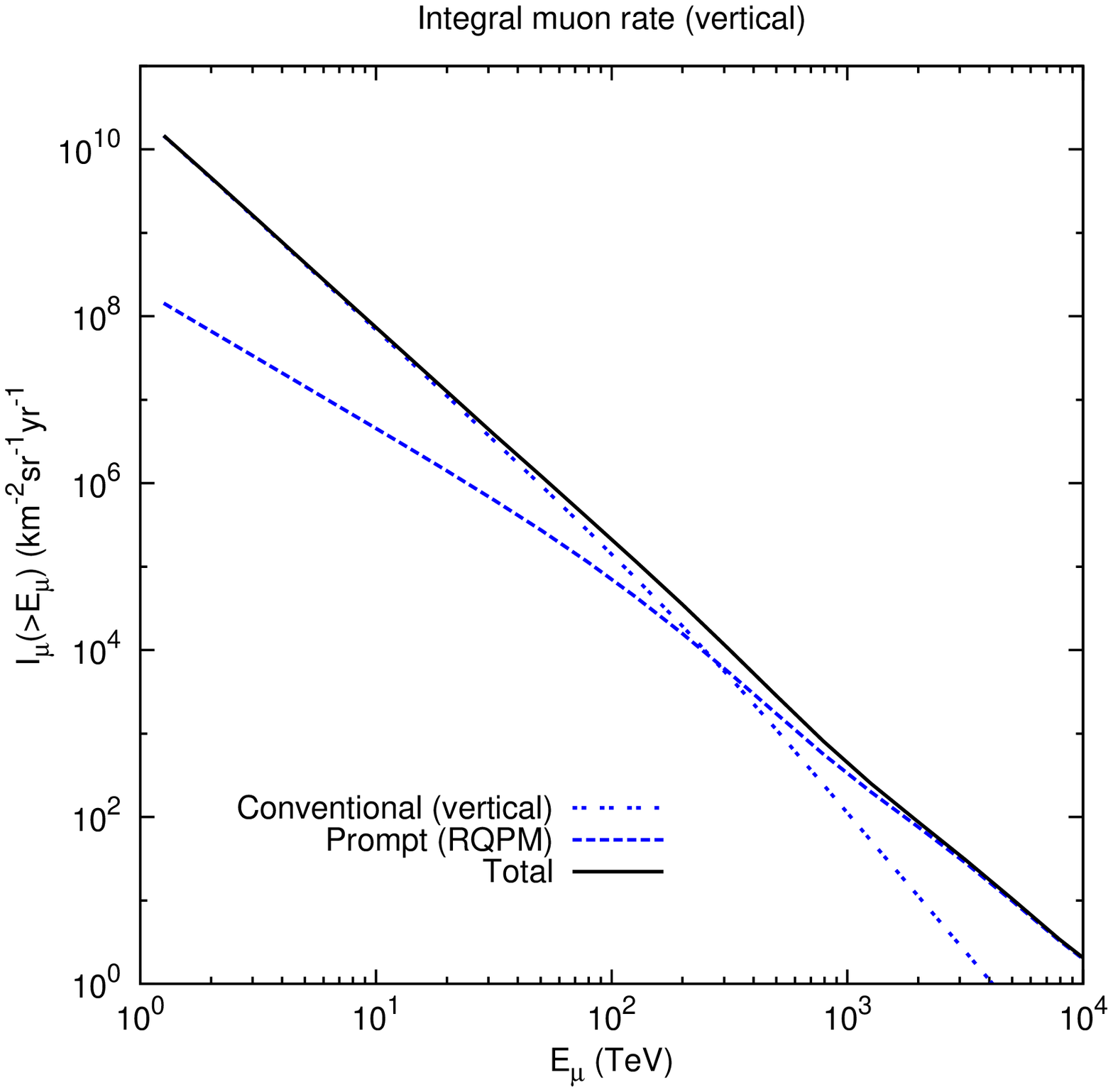}
\caption{Integral muon rate.  Left: with rescaled ERS model for prompt muons; Right: with rescaled
RQPM model for charm.} 
\label{fig5}       
\end{figure*}

\section{Expected rates}
\label{sec5}
The fluxes described above can be used to estimate the rate of events in a kilometer scale detector.
For atmospheric muons the rate per year is simply the flux multiplied by $3\times 10^7$ seconds/yr
and by $10^{10}$~cm$^2$/km$^2$.  The corresponding integral
rate of events $I_\mu(>E_\mu)$ is shown in Fig.~\ref{fig5}.  

\begin{figure*}
\centering
\includegraphics[width=7.5cm]{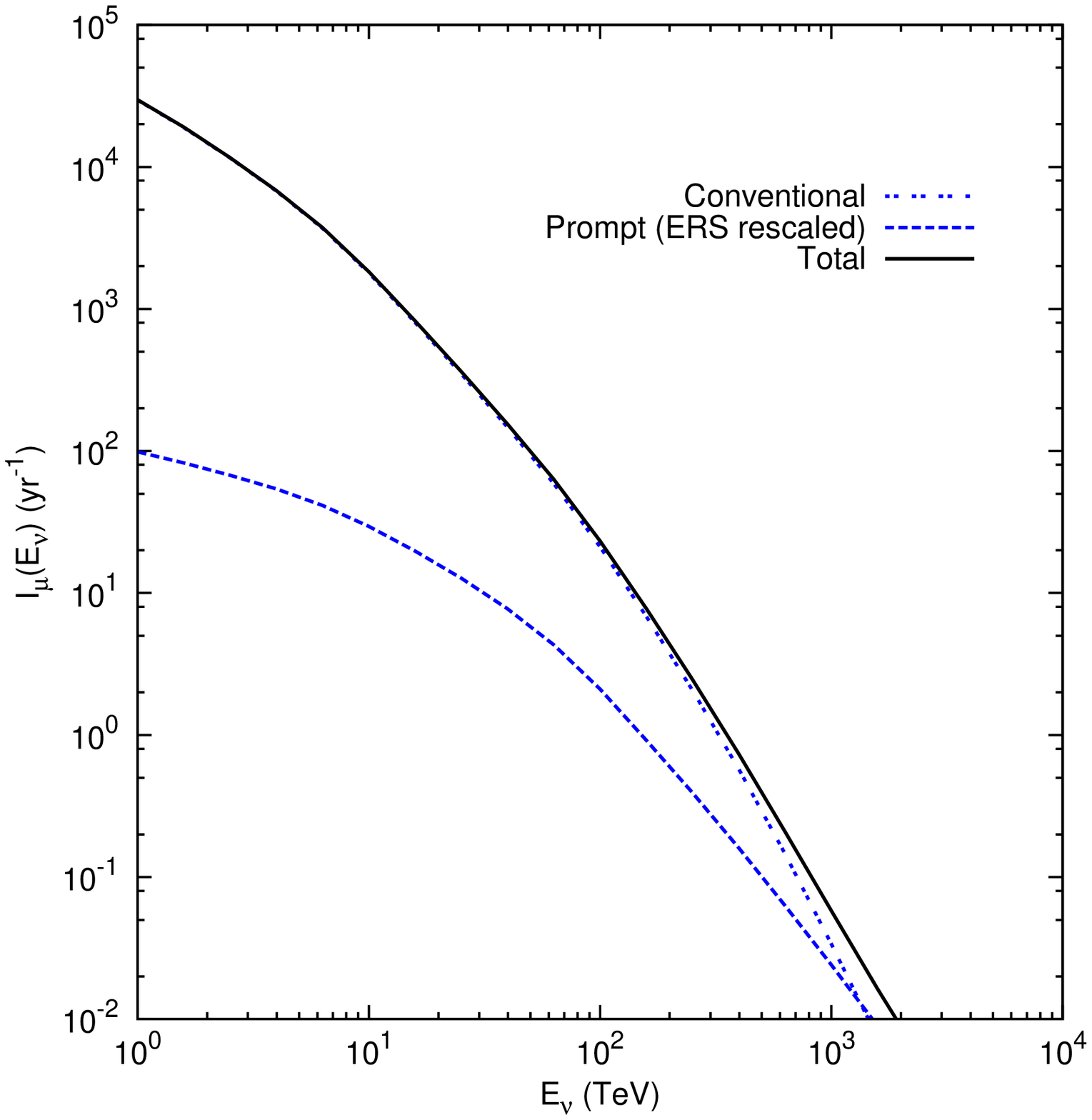}\,\includegraphics[width=7.5cm]{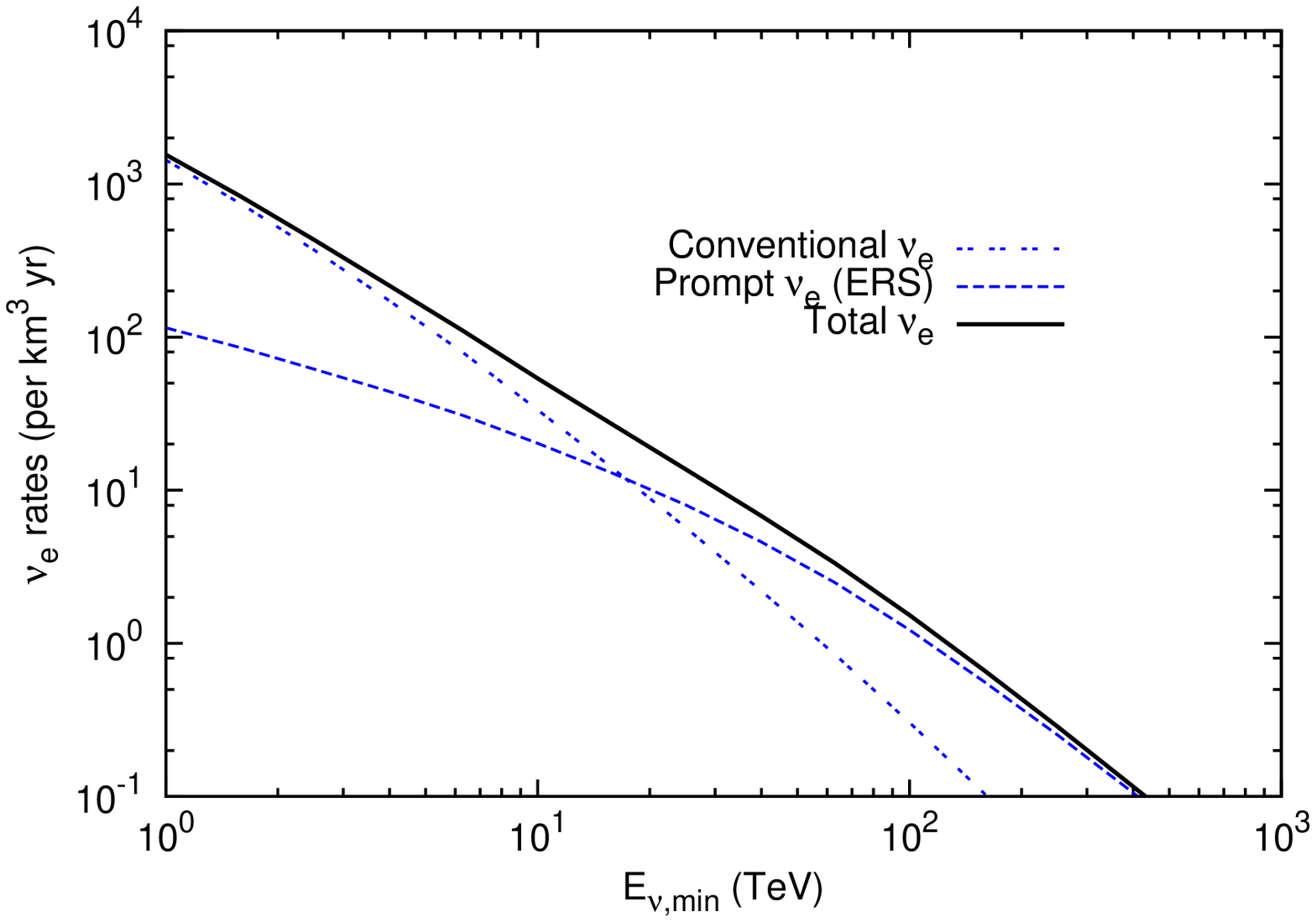}
\caption{Estimate of the rate of atmospheric neutrino
interactions per year in a km$3$ detector.  Left: neutrino-induced muons per km$^2$ and with
zenith angles from $90^\circ$ to $120^\circ$; 
Right: electron neutrinos
with vertices inside 1 km$^3$.
ERS is rescaled in both plots, as discussed in the text.}
\label{fig6}       
\end{figure*}

The rate of neutrino-induced muons can be obtained in a similar way, with one additional step.
It is necessary to calculate the effective area to convert the rate of neutrinos with
trajectories passing through the detector to a rate of neutrino-induced muons.  Effective area
is defined in such a way that $\phi(E_\nu,\theta)\times A_{\rm eff}(E_\nu,\theta)$ is the rate of neutrino-induced
muons per second per sr at zenith angle $\theta$.  Explicitly
\begin{eqnarray}
A_{\rm eff}(E_\nu,\theta)&=&\epsilon(E_{\rm th},\theta)\,
A(\theta)\,P_\nu(E_\nu,E_{\rm th})\\ \nonumber
&\times &\exp\{-\sigma_\nu(E_\nu)N_AX(\theta)\},
\label{EqAeff}
\end{eqnarray}
where $P(E_\nu)$ is the probability that a neutrino converts
and produces a muon that reaches the detector with enough energy to be reconstructed.
Absorption of neutrinos in the Earth becomes significant in the $10$ to $1000$ TeV range,
first for vertically upward trajectories and for neutrinos with zenith
angles $\sim120^\circ$ around a PeV.
An accurate calculation of $A_{\rm eff}$ requires a detector simulation.  Here I use
an estimate for an ideal km$^2$ detector from Ref.~\cite{TG-Tokyo08} and estimate the
rate of neutrino-induced muons in the zenith angle range from horizontal to $-120^\circ$.
The result is shown for the ERS assumption in the left panel of Fig.~\ref{fig6}.

Electron neutrinos must interact in the detector to be identified as cascades in the
detector.  Such cascades are virtually indistinguishable from neutral current
interactions of muon or electron neutrinos of energy $E_\nu\sim E_{\nu_e}/y$,
where $y$ is the inelasticity of the neutral current neutrino interaction.  The neutral
current interactions of atmospheric $\nu_\mu$ make a comparable contribution
to cascades for the conventional atmospheric neutrinos because the flux of $\nu_\mu$
is significantly higher than that of $\nu_e$.  For the charm component, however,
the neutral current contribution is relatively unimportant because of the equality
of the fluxes of prompt $\nu_e$ and $\nu_\mu$.  
The integral rate of $\nu_e$ interactions from all directions is 
\begin{equation}
R(>E_\nu)\,=\,4\pi N \times T\times \int_{E_\nu} \,\sigma_{\rm cc} \phi(E_\nu){\rm d}E_\nu,
\label{nue-rate}
\end{equation}
where $\phi(E_\nu)$ is the spectrum of $\nu_e+\bar{\nu}_e$ averaged over all directions
and $\sigma_{\rm cc}$ is the charged current cross section, taken here from Ref.~\cite{CooperSarkar}.
N is the number of nucleons per cubic kilometer of ice and T=1 year.  The resulting estimate of
the rate of atmospheric $\nu_e$ interactions is shown in the right panel of Fig.~\ref{fig6}.
The plot includes the effect of neutrino shadowing by the Earth in the upward hemisphere~\cite{Gandhi}.
This amounts to a suppression for the whole sky of 9\% at 100 TeV and 23\% at a PeV.
The plot shows about 5 electron neutrino interactions above 100 TeV per km$^3$ of ice per year
assuming ideal (full) efficiency.  The spectrum is steeply falling so that less than
one such event in ten years is expected above a PeV.

\section{Summary comments}
\label{sec6}
In view of the recent observation by IceCube of two cascade-like events with observed energy
just above a PeV~\cite{Aya}, together with the observation of several high=energy cascades
reported at this meeting~\cite{Middell}, it is important to achieve a good understanding
of the background from atmospheric neutrinos in the energy region around 100 TeV and above.
The work outlined in this paper is just a start.

There are large uncertainties in the primary spectrum in the TeV range and above that
need to be assessed.  The model used here is just one possibility.
A model for charm production that includes recent data from LHC~\cite{ALICE,LHCb} in its fits
needs to be developed.  As an example, the framework for including charm within
SIBYLL~\cite{SIBYLL} exists~\cite{SIBcharm}, but the parameters need to be
tuned to data over a wide range of energy and phase space.
An updated model could then be used for a full Monte Carlo
simulation including charm.  The analytic approximations will remain an important tool
to supplement the Monte Carlo, for example to track the consequences of uncertainties
in the input spectrum and hadronic interactions for the expected fluxes~\cite{Agrawal},
as well as to parameterize and extrapolate the limited statistics of the full 
Monte Carlo~\cite{FBD}.

\section*{Acknowledgments}
I am grateful to colleagues in the IceCube collaboration for discussions in various
contexts that stimulated this work.  My research related to IceCube is supported in part
by NSF-PHY-1205809.  My research related to hadronic interaction models is supported
in part by the U.S. Department of Energy, DE-SC0007893.

\end{document}